\documentclass[]{spie}  

\usepackage{amsmath,amsfonts,amssymb}
\usepackage{graphicx}
\usepackage[colorlinks=true, allcolors=blue]{hyperref}

\title{CHARA Array adaptive optics: complex operational software and performance}

\author[a,*]{Narsireddy Anugu}
\author[b]{Theo ten Brummelaar}
\author[b]{Nils H. Turner}
\author[b]{Matthew D. Anderson}
\author[c]{Jean-Baptiste Le~Bouquin}
\author[b]{Judit Sturmann}
\author[b]{Laszlo Sturmann}
\author[b]{Chris Farrington}
\author[b]{Norm Vargas}
\author[b]{Olli Majoinen}
\author[d]{Michael J. Ireland}
\author[e]{John D. Monnier}
\author[f]{Denis Mourard}
\author[b]{Gail Schaefer}
\author[b]{Douglas R. Gies}
\author[g]{Stephen T. Ridgway}
\author[h]{Stefan Kraus}
\author[i]{Cyril Petit}
\author[f]{Michel Tallon}
\author[i]{Caroline B. Lim}
\author[f]{Philippe Berio}

\affil[a]{Steward Observatory, Department of Astronomy, University of Arizona, Tucson, USA}
\affil[b]{CHARA Array, Georgia State University, Atlanta, GA 30302, USA}
\affil[c]{Institut de Planetologie et d'Astrophysique de Grenoble, Grenoble 38058, France}
\affil[d]{The Australian National University, Canberra, ACT 2600 Australia}
\affil[e]{University of Michigan, Ann Arbor, MI 48109, USA}
\affil[f]{UCA/OCA/CNRS, Campus Valrose, 28 avenue Valrose, 06108 Nice France}
\affil[g]{NOIRLab, NSF's National Optical-Infrared Astronomy Research Laboratory, 950 N. Cherry Ave., Tucson, AZ 85719, USA}
\affil[h]{School of Physics and Astronomy, University of Exeter,  Exeter, Stocker Road, EX4 4QL, UK}
\affil[i]{DOTA, ONERA, Université Paris Saclay, F-91123 Palaiseau - France}

\authorinfo{* Steward Observatory Fellow in Astronomical Instrumentation and
Technology.  \\ E-mail: \href{nanugu@arizona.edu}{nanugu@arizona.edu}}

\begin{document} 
\maketitle

\begin{abstract}
The CHARA Array is the longest baseline optical interferometer in the world. Operated with natural seeing, it has delivered landmark sub-milliarcsecond results in the areas of stellar imaging, binaries, and stellar diameters. However, to achieve ambitious observations of faint targets such as young stellar objects and active galactic nuclei, higher sensitivity is required.  For that purpose, adaptive optics are developed to correct atmospheric turbulence and non-common path aberrations between each telescope and the beam combiner lab. This paper describes the AO software and its integration into the CHARA system. We also report initial on-sky tests that demonstrate an increase of scientific throughput by sensitivity gain and by extending useful observing time in worse seeing conditions. Our 6 telescopes and 12 AO systems with tens of critical alignments and control loops pose challenges in operation. We describe our methods enabling a single scientist to operate the entire system.
\end{abstract}


\section{INTRODUCTION}
\label{sec:intro}  

The CHARA Array\cite{tenBrummelaar2005,tenBrummelaar2016SPIE.9907E..03T,Schaefer2020} is the world longest baseline optical or near-infrared interferometer with six 1-meter diameter telescopes and baseline up to $B=331$\,m delivering sub-millisecond angular resolution (e.g., $\lambda/2B\sim0.6$mas at H-band). The six telescopes of the CHARA Array are arranged in a Y-shaped configuration, and they provide 15 interferometric baselines and 20 closure phases.  The CHARA Array has been a very successful observatory so far by delivering landmark results in the areas of stellar imaging\cite{Monnier2007, Roettenbacher2016}, binaries \cite{Kloppenborg2010, Kraus2020Sci...369.1233K}, expansion phase of a nova explosion \cite{Schaefer2014},  and stellar radii measurements\cite{Boyajian2012ApJ...757..112B} with a sub-millisecond angular resolution.  However, to achieve the challenging goals of imaging faint targets such as young stellar objects and active galactic nuclei, and measuring radii of TESS exoplanet host targets require higher instrument sensitivity. 

Since obtaining the first-fringes, the beam combiner laboratory of the CHARA Array is equipped with fast tip-tilt systems\cite{Sturmann2006SPIE.6268E..3TS}, although the original goal was to deploy adaptive optics systems (AO) for the telescopes and in the laboratory. Since 2012, the CHARA Array adaptive optics programs are under development in the beam combiner laboratory and at the telescopes. They are developed in two phases, accommodating funding realities (PI: Theo ten Brummelaar).  These adaptive systems enable higher throughput by increasing the Strehl ratio of each beam.  The light collected by the telescopes is injected into single-mode fibers, is a standard practice currently in the field, for spatial filtering\cite{Shaklan1987} before the beam combination inside a beam combiner (e.g., MIRC-X\cite{Anugu2018a,Anugu2020AJ....160..158A, Anugu2020}, MYSTIC\cite{Monnier2018} and SPICA\cite{Mourard2017, Pannetier2020}). The sensitivity of optical interferometers is an order of a magnitude lower than those in most classic telescopes is due to: 
\begin{itemize}
    \item Too many mirrors in the light path from the sky to a beam combiner, e.g., the CHARA Array contains 21 mirrors. Furthermore, telescope and instrument vibrations play a huge role.
    \item Non-common-path errors in an interferometer are orders of magnitude greater than those in the most telescope AO systems.
    \item Interferometers combine coherent light to make fringes, but because of atmospheric turbulence, only short coherent integrations are allowed -- smaller than the atmospheric coherence time, typically 10-25ms.
    \item Atmospheric turbulence severely limits the light coupling into the single-mode fibers with a distorted point spread function\cite{Shaklan1987}. For a pupil with a central obstruction of 25\% of the pupil diameter, the expected coupling efficiency into a single-mode fiber is 0.69 times the Strehl ratio\cite{Ruilier1998SPIE.3350..319R}. To improve the instrument throughput, we need high Strehl ratio beams at the tip of single-mode fibers. 
\end{itemize} 

Without adaptive optics, the sensitivity of observing the faintest star with any beam combiner at the CHARA Array is lower than eight magnitudes in H or K-bands. 

The CHARA Array has twelve adaptive optics systems in total : 
\begin{itemize}
\item \textbf{Tel-AO (in Phase II)}\cite{ten2018SPIE10703E..04T,ten2016AAS...22742702T,Che2013JAI.....240007C,ten2012SPIE.8447E..3IT} Six at the telescope sites to correct the tip-tilt and high-order aberrations induced by the atmospheric turbulence.  These AO systems provide high Strehl ratio beams to the vacuum tubes, which transport the light to the beam combiner lab. 
\item \textbf{Lab-AO (in Phase I)}\cite{ten2014SPIE.9148E..4QT, Che2014SPIE.9148E..30C} Six at the beam combiner lab to correct significant non-common path aberrations that arose between the telescope and the beam combiner instruments.  The non-common path aberrations errors are due to several hundred meters of the light path in between the telescope and the beam combiner, including several mirrors and delay lines. 
\end{itemize}

The Lab-AO systems were commissioned already in Phase~I\cite{ten2014SPIE.9148E..4QT} and are in operation since 2014 for science observations as reported in the previous SPIE proceedings\cite{ten2014SPIE.9148E..4QT, Che2014SPIE.9148E..30C, Che2013JAI.....240007C,ten2012SPIE.8447E..3IT}. The Tel-AO systems have been commissioned since 2018, in incremental steps, one telescope after another. The preliminary optical design of the Tel-AO systems is presented in 2018 SPIE proceedings\cite{ten2018SPIE10703E..04T}. In this presentation, we report the software architecture, operation, and results. 

A few challenges we face in the operation of the twelve adaptive optics systems are:

\begin{itemize}
\item \textbf{Obtain high-Strehl ratios by making a good AO correction} For this, the software development requires acquiring low-latency frames from the wavefront sensor camera, computation of slopes, computation of control matrix, correction of high-order aberrations with the ALPAO deformable mirrors (DM), and tip-tilts with the telescope M2.

\item \textbf{Alignment between Tel-AO and Lab-AO} The Lab-AO is connected to the Tel-AO using a metrology system. The Lab-AO wavefront sensor (WFS)  records beacon light sent from the telescope that uses a wavelength outside that of the science band of interest to measure the non-common path aberrations\cite{ten2018SPIE10703E..04T}. A dichroic at the telescope reflects some visible wavelengths of starlight into the Tel-AO WFS, and it also sends the beacon light into the Lab-AO WFS. The beacon light must follow the same light path as the star. We require an automatic alignment in pupil and focus.

\item \textbf{Complexity in operation} Control of six telescopes and twelve adaptive optics systems by an operator is a challenge. Several Graphical User Interface (GUI)s execute the control of these systems.  Each system has its own GUI, for instance, GUI for the telescope, acquisition camera, telescope tip-tilt, Tel-AO, and Lab-AO. Many GUIs invite inefficiency and error. Our challenge is to present a manageable interface for the operator. For example, the Tel-AO and Lab-AO systems alone require six monitors to fit their GUIs. For each slew to a new star target, the sequence includes the telescope pointing and tracking, star acquisition with the acquisition camera, locking the closed-loops of the tip-tilt, Tel-AO, and Lab-AO. We required an efficient software system and an integrated, stable, and easy operable GUI for each telescope to do all these steps in a small amount of time to maximize the time for the fringe finding and recording. 
\end{itemize}

\section{Overview of CHARA adaptive optics}
\begin{figure} [h!]
\centering
\includegraphics[width=\textwidth]{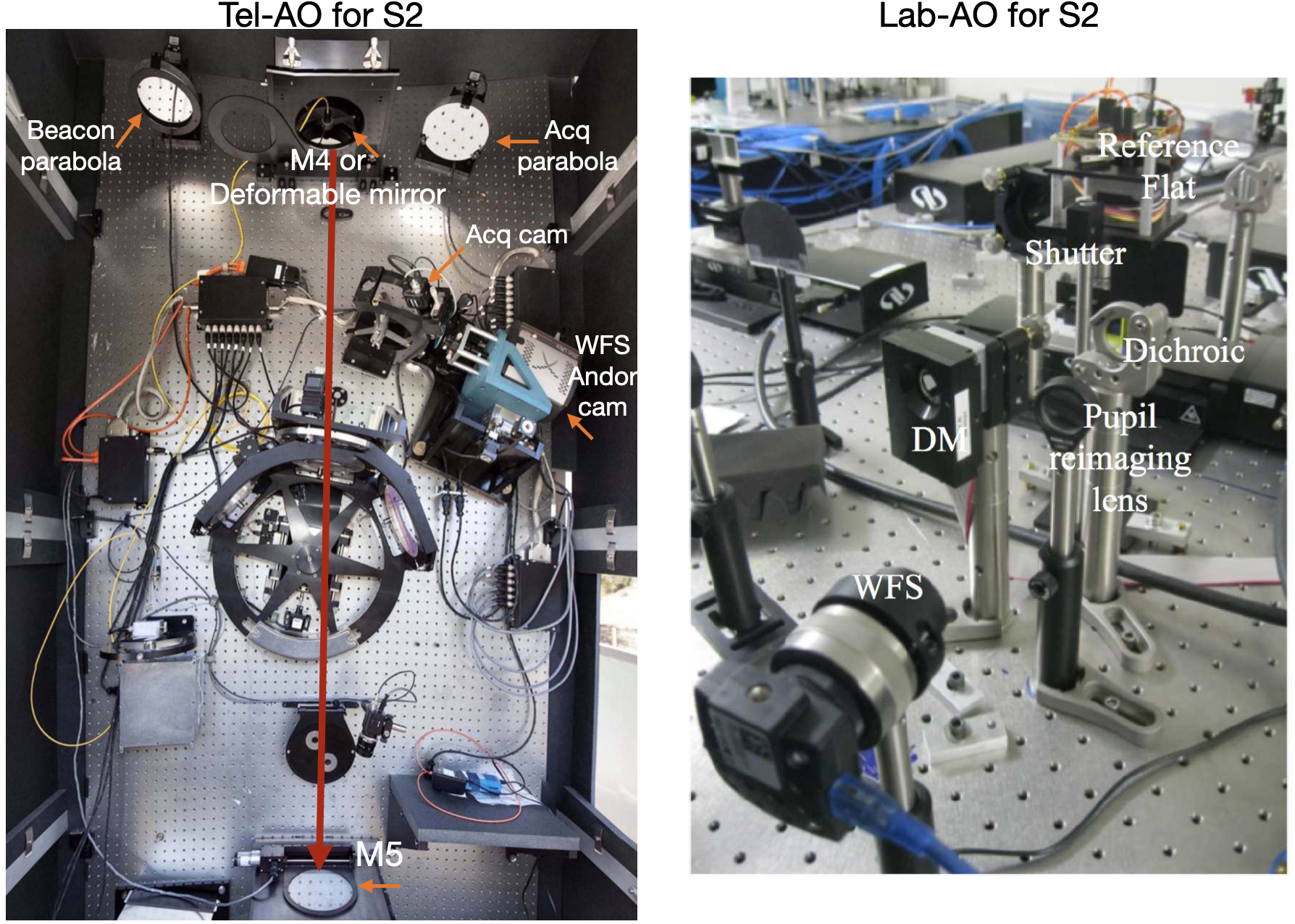}
\caption[example]{ \label{fig:ao-as-built}  Photographs of the Tel-AO (left) and Lab-AO systems (right) as built. }
\end{figure}

Figure~\ref{fig:ao-as-built} shows the Tel-AO and Lab-AO systems as built. A brief description of the optical design of each system is followed below.

\subsection{The telescope adaptive optics system (Tel-AO) }
The Tel-AO is designed to correct the atmospheric turbulence above the telescope and improve the Strehl ratio of the beams in H or higher wavelengths. Each Tel-AO system consists of (a) a $7\times7$ lenslet based Shack-Hartmann wavefront sensor and a 500~Hz fast frame rate and sub-electron readout noise, Andor 897 EMCCD camera\cite{Che2013JAI.....240007C} and (b) an ALPAO voice-coil based deformable mirror with 61 actuators\cite{ten2018SPIE10703E..04T}. The Tel-AO DMs are placed in the position of the M4 mirror to limit new reflections to the beam train. This mirror position is at 45 degrees, and this orientation results in an ellipse pattern pupil. The DM and the Shack-Hartman match this ellipse shape. The Tel-AO has three dichroic glasses: BARE, VIS, and YSO. Based on a science requirement, an appropriate dichroic filter is chosen to maximize flux into the Tel-AO WFS. Table~\ref{tab:dichroic} presents the dichroic filter and their percentage of share into the Tel-AO WFS.

The expected performance of Tel-AO is (i) more than one magnitude sensitivity improvement in good seeing conditions (Fried parameter $r_0>15$cm at $\lambda=0.5\mu$m), and (ii) 1-3 magnitude sensitivity improvement for  $5<r_0<15$cm and that means three times of more high quality observing time. 

\begin{table}[ht]
\caption{Tel-AO dichroic filter selection and \% light shared with the Tel-AO WFS}
\label{tab:dichroic}
\begin{center}       
\begin{tabular}{cc}
\hline
Dichroic glass & Share to the Tel-AO WFS \\
\hline
BARE & 4\% \\ 
VIS  & 20\% \\
YSO  & 100\% \\
\hline
\end{tabular}
\end{center}
\end{table}

\begin{table}[ht]
\caption{Deformable mirror specifications} 
\label{tab:dm_spec}
\begin{center}       
\begin{tabular}{lll}
\hline
Parameter & Tel-AO & Lab-AO  \\
\hline
DM actuators & ALPAO 61  &  OKO MMDM 37  \\
Size &  18cm  & 15 mm\\
Dynamic range & 16$\mu$m &  9$\mu$m \\
Inter-actuator stroke & $4\mu$m & 0.5$\mu$m \\
First resonance & 500 Hz  & - \\
Settling time & 2ms &  - \\
Mirror best flat & $<30$~nm & 400~nm\\
Optimized for H and K-bands  & Spacing of sub-apertures is $\sim14$cm on telescope pupil & \\
&  $r_0\sim5$cm at $0.5\mu$ m, $r_0=21$cm at 1.6$\mu$m  & \\
\hline
\end{tabular}
\end{center}
\end{table}

\subsection{The laboratory adaptive optics system (Lab-AO)}
The Lab-AO systems are placed after the delay lines in the beam combiner laboratory and just before the beam combiner instruments\cite{ten2014SPIE.9148E..4QT}. The Lab-AO is intended to correct for non-common path errors in the system, which are significant as there are several hundred meters of the path in each beam, each including several mirrors and delay line carts.

To reduce the cost of the Lab-AO systems, use off-the-shelf DM (OKO MMDM with 37 actuators) and $6\times6$ lenslet imaged with a USB CCD camera. Since the Lab-AO uses the beacon from the telescope, there is enough light, and a high-readout noise camera is not a problem. The performance of the Lab-AO systems are not at the level of a standard adaptive optics system (see Table~\ref{tab:dm_spec} and \ref{tab:wfs_spec}), and as it is used only for quasi-static aberrations: (i) it has a lower number of actuators, and its WFS operates at a lower speed~100Hz, and (ii) The measured wavefront errors are in a low order compared to the Tel-AO WFS estimated wavefront errors.

\begin{table}[ht]
\caption{Wavefront sensor specifications}
\label{tab:wfs_spec}
\begin{center}       
\begin{tabular}{lll}
\hline
Parameter & Tel-AO & Lab-AO \\
\hline
Lenslet sub-apertures & $7\times7$ & $6\times6$\\
Effective sub-apertures & 36 & 32 \\
Sub-aperture field of view & $6.7^{\prime\prime}$ & $4.6^{\prime\prime}$ \\
Pixel per sub & 9 & 5 \\
camera window size & $90\times90$ & $160\times128$ \\
Frame rate & 500Hz & 100Hz \\ 
Sensitivity R-band & $<14$ with EMCCD gain & $<4$\\
Zernikes measured & 21 & 21\\
\hline
\end{tabular}
\end{center}
\end{table}

\section{Control software architecture and development}

\subsection{The Tel-AO computer hardware}
The computer server plays a critical role in the adaptive system's performance. It needs to read wavefront sensor images with a low-latency and process them for wavefront slopes and compute the command matrix without much additional delay.  The Tel-AO computer hardware is entirely commercial off-the-shelf (COTS), utilizing standard components. There are six computers, one for each telescope, and the fundamentals of these computers and data acquisition frame grabbers are identical, meaning they use the same motherboard, processor, and RAM. The computers' selection is driven by low-power consumption as the WFS camera is placed inside the telescope dome. The Andor 897 EMCCD works with USB control with limited length cables.  The computers are selected with the specification of Intel CPU 6 Cores with i7-8700 clock speed 3.20GHz. These computers are operated with the Ubuntu Linux low latency operating systems with kernel 4.4.0. 

\begin{figure} [h!]
\centering
\includegraphics[width=0.8\textwidth]{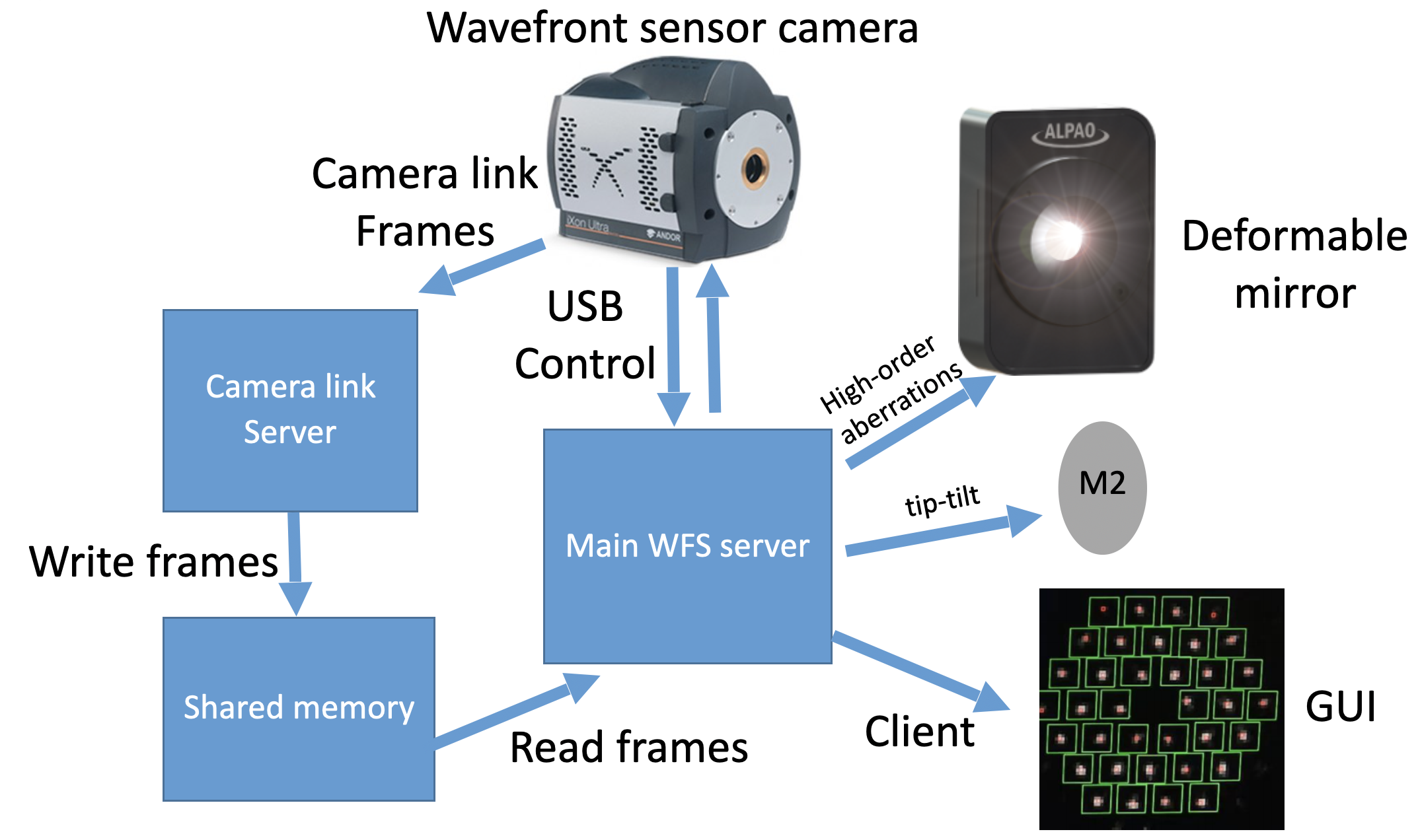}
\caption[example]{ \label{fig:telao_software} 
The Tel-AO software layout. The camera link server reads the frames from the WFS Andor EMCCD camera through BitFlow Neon frame grabber and writes them to an instrument shared memory in a circular buffer so that the main WFS main server uses them. The WFS  main server reads the frames from the instrument shared memory and calculates the wavefront slopes from each new science frame. Next, the DM commands are computed by multiplying the slopes matrix with the slopes-to-actuator reconstructor matrix. The tip/tilt is computed by taking the average of slopes and then sent to telescope M2 for correction with appropriate gain. The rest higher-order corrections are sent to the ALPAO deformable mirror. The wavefront sensor images and deformable mirror shapes are sent to the client GUI for monitoring and operation.}
\end{figure} 

\subsection{The Tel-AO low latency software}
The software development of the Tel-AO systems (see Figure~\ref{fig:telao_software}) involves acquiring low-latency wavefront sensor images, computation of slopes\cite{anugu2018peak}, computation of control matrix, correction with the ALPAO deformable mirrors, and telemetry data recording. Furthermore, the software handles the automated stabilization of pupil and beacon drifts. We follow a modular design, whereby multiple processes/threads interact through shared memory to process the camera data at a full-frame rate with minimum latency.

\begin{figure} [ht]
\centering
\includegraphics[width=0.7\textwidth]{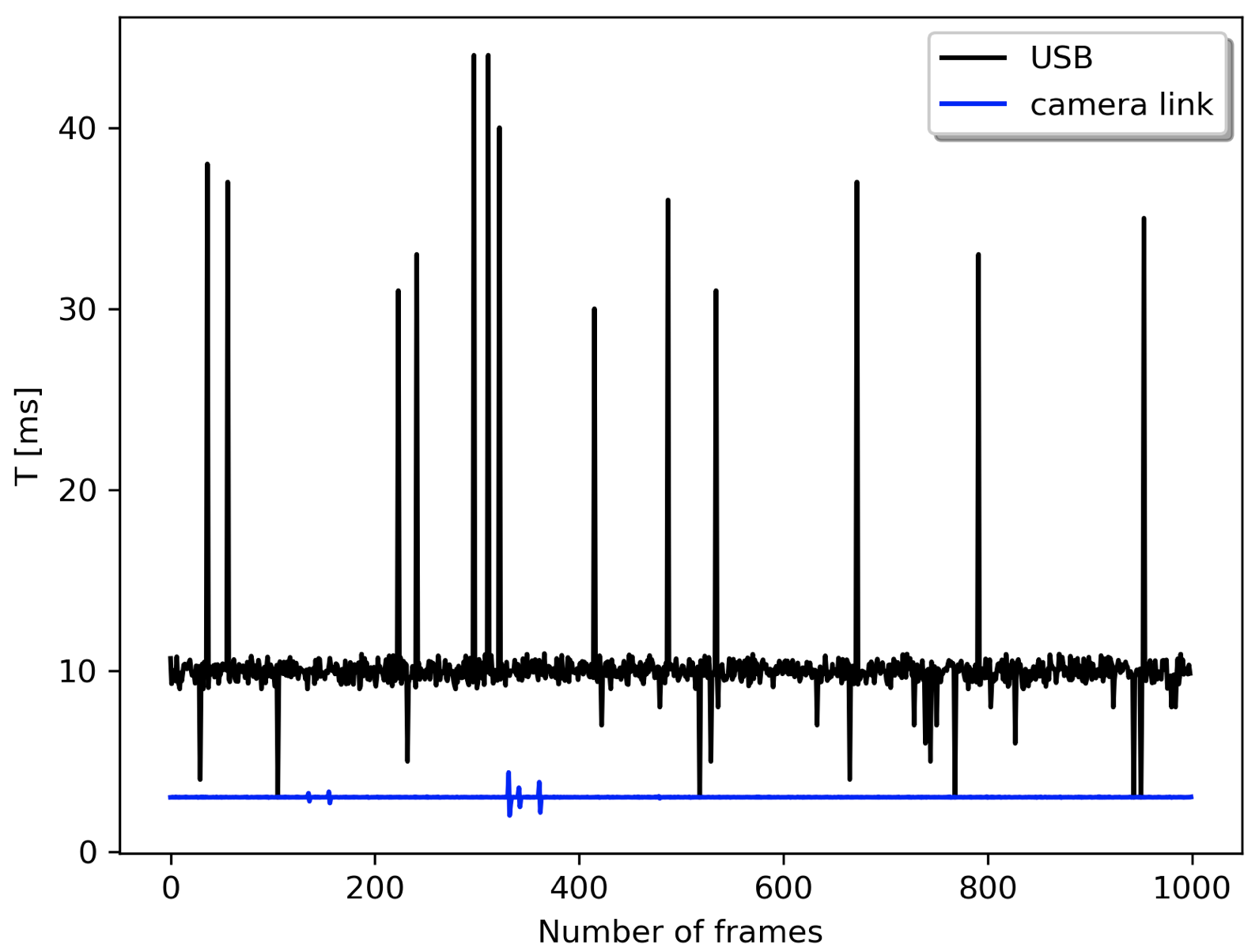}
\caption[example]{ \label{fig:usb_mising_frames} 
USB data data acquisition of frames (black color) in a comparison  to the camera-link data acquisition (blue color). The integration is 3ms but the data points are the time difference between two consecutive frame grabs. The USB acquisition has a lot of jitters, in a worst-case, a grab gets 8 frames in a chunk (i.e., 8 frames of delay). In the case of camera-link, there is also jitter but very small. }
\end{figure}

We realized the default USB-based data acquisition of the Andor camera is missing frames (see Figure~\ref{fig:usb_mising_frames}), and as a solution, we implemented a low latency frame grabbing solution with a BitFlow Neon CLB frame grabber. Before trying BitFlow frame grabber, we tried Matrox Radient eV-CL frame grabber, and it did not work as the Andor camera link out is non-standard in the market: base configuration, 3-tap interface, and 16- bit greyscale. We were able to find another working frame grabber, Epix, but the latency jitter was high compared to the BitFlow Neon CLB frame grabber. The frames produced by the camera are written to an instrument shared memory, and those are accessed directly by the main wavefront sensor server. The main server computes slopes from raw images, commands from slopes matrix, and controls the deformable mirror for the correction of high-order aberrations and the telescope M2 mirror for the correction tip-tilts  (see Figure~\ref{fig:telao_software}).

In Figure~\ref{fig:usb_mising_frames}, we compare the USB based data acquisition and camera-link based data acquisition.  The time difference between two consecutive frame grabs is recorded 1000 times. In a low latency situation, one expects to get one frame for each frame grab request, and the time difference between two consecutive grabs is more or less equal. In a jitter frame grabbing situation, we get more than one frame in a grab, and the time difference between two consecutive grabs is not the same (sometimes considerable big time and some less time).  The USB data acquisition has many jitters in frame arrival, and the average consecutive frame arrival is for every 10ms for an integration time of 3ms. In the worst case, a USB frame grab gets eight frames in a chunk (i.e., eight frames of delay, which means a delay between consecutive frames is more significant than 24ms). In the camera-link case, there is jitter in the frame arrival but very small -- average consecutive frames arrival for every 3.1ms for the integration of 3ms. Here we do not report the pure delay.

We list below the Andor functions used in our software. These were critical in getting a low-latency arrival of frames from the Andor camera; otherwise, we noticed an additional pure delay of two frames, at 2ms integration. The measure of pure delay of the WFS Andor camera is attempted by sending a series of impulse pokes on the ALPAO deformable mirror and noticing the resulting shapes on the wavefront sensor, and computing the time delay. The known deformable mirror delay is subtracted from this measurement. For a 2ms integration time, we measure 1.8ms pure delay, which is the sum of the camera frame transfer time and readout time. For comparison, this is almost similar to NAOMI\cite{Woillez2019A&A...629A..41W}, and their reported value is 1.54ms. They use the same Andor 897 camera, but the difference of lower 0.26ms maybe due to the usage of their hard real-time operating machine for frame grabbing than our COTS machine.

\begin{verbatim}
vspeed(2)                               -> 0.5 microseconds
hspeed(2)                               -> 10MHz
SetFrameTransferMode(1)                 -> readout in Frame Transfer Mode
SetReadMode(4)                          -> readout mode image
SetIsolatedCropMode(1, 90, 90, 1, 1)    -> read cropped 90x90 pixels with 
                                           starting vertical and horizontal bins
SetIsolatedCropModeType(1)              -> low latency mode saves 2 frames of delay
SetAcquisitionMode(5)                   -> acquisition run till abort
StartAcquisition()                      -> starts the acquisition 
\end{verbatim}

The influence functions of all DM actuators were measured in a standard method using the wavefront sensor. Each actuator is "poked" and the resulting influence on the wavefront is measured. The actuator poking was done in positive and negative directions to minimize the measurement's noise by taking the average of both amplitudes. Furthermore, 100 wavefront measurements were taken for each poke to average out and reduce the noise of the WFS detector. 

A challenge for telemetry data recording is management since we have twelve systems.  We only save telemetry data when requested by the user as all the AO systems generate more than 1TB of data per night. An important feature of our telemetry data recording and saving is that the data is recorded into memory and saved to disk without interrupting priority tasks. The telemetry data includes time-stamped wavefront sensor images, measured slopes, aberrations, and actuator-to-slopes reconstructor matrix. This telemetry data used to estimate the performance of the AO system, and as well as it is used to correlate with science fringe data during post-processing.


\begin{figure} [h!]
\centering
\includegraphics[width=\textwidth]{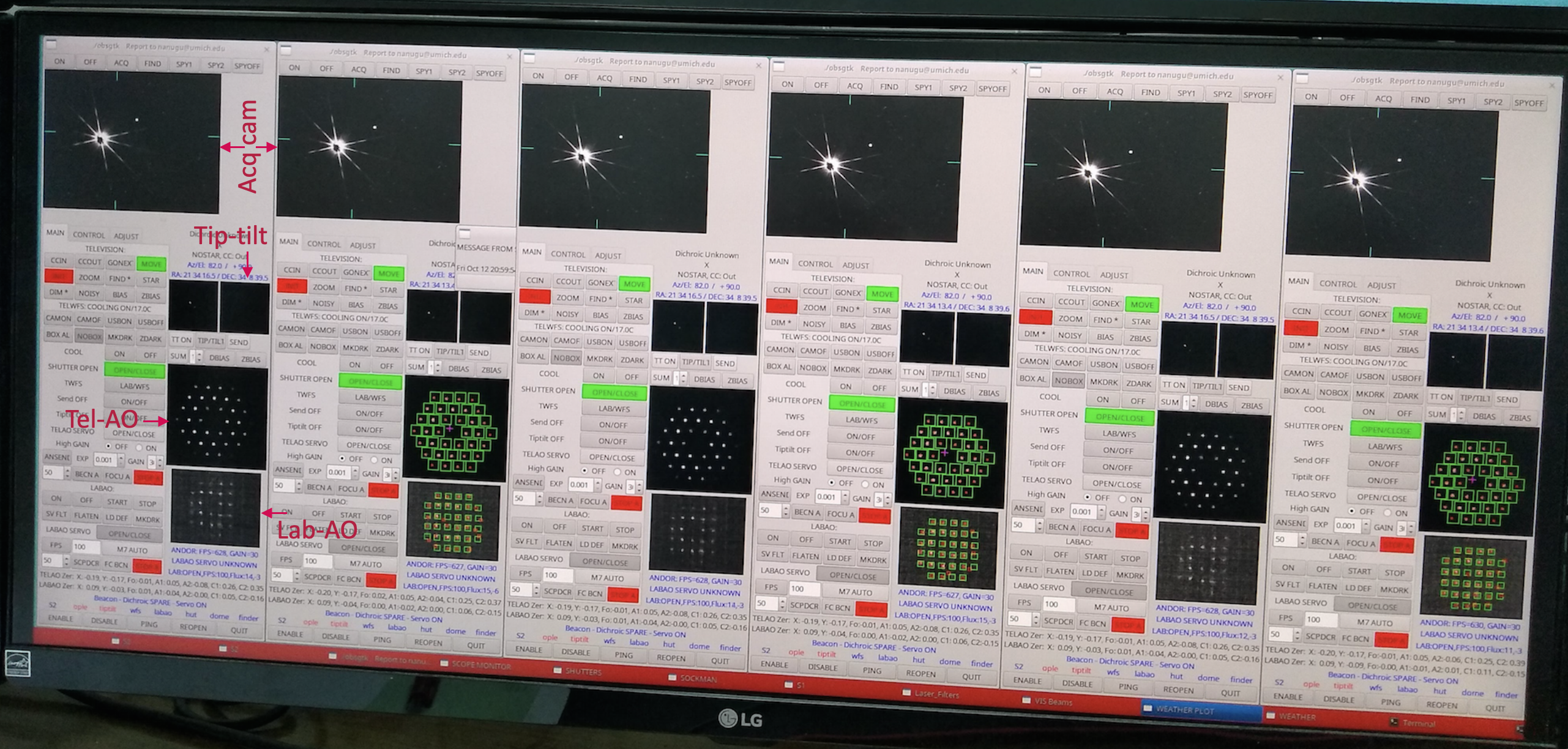}
\caption[example]{ \label{fig:obsgtk} 
A photograph of the main observing GUI at the CHARA Array, \texttt{obsgtk}. Each  \texttt{obsgtk} controls one telescope and has all the controls for the telescope and its acquisition camera, tip-tilt, Tel-AO, and Lab-AO, including the real-time displays of the acquisition camera and wavefront sensors of Tel-AO and Lab-AO. This GUI is optimized for screen space -- all six GUIs fit in a single screen monitor, time-efficient usage, and stability by keeping only the required observing controls together. This photograph is taken with telescope beacon was ON.}
\end{figure} 

\subsection{Observational complexity of operating six Tel-AO + six Lab-AO}
The adaptive optics systems software is built leveraging on the CHARA Array client/server architecture\cite{tenBrummelaar2005}.  Servers publish the information, and appropriate clients subscribe to the servers to get the information. We do not have a central database system. The hardware (cameras, DMs, and other actuators) is controlled by the servers installed at the site of AO systems. We use GNU Tool Kit (GTK) Graphical User Interface (GUI) clients installed on the computers at the control room to send user configurations to the servers and monitor the real-time display of acquisition cameras and wavefront sensor cameras.  

The control of six telescopes, six acquisition cameras, twelve adaptive optics systems, and six tip-tilt systems is challenging for an operator considering several GUIs. The GUIs of twelve adaptive optics systems take a lot of computer screen space and would require around six monitors. Requiring hands-on interaction with a large number of subsystem GUI’s is not a practical solution for a telescope facility that must work with a single operator. A software solution is required that simplifies the control problem.  This problem is approached with an integrated GUI for each telescope for saving screen space, operational efficiency, time-saving, and easy usability. We took critical real-time displays, buttons, and features from the engineering GUIs of different systems and put them into the integrated GUI. In the end, the GUIs of all the telescopes, acquisition cameras, adaptive optics systems, and tip-tilts systems are fit into a single computer display monitor, as shown in Figure~\ref{fig:obsgtk}. 

The integrated GUI, \texttt{obsgtk}, connects and communicates with several servers including telescope (e.g., \texttt{S1}), acquisition camera (e.g., \texttt{S1\_ACQ}), dome (e.g., \texttt{dome\_S1}), motors server (e.g., \texttt{S1\_HUT}), Tel-AO (e.g., \texttt{wfs\_S1}) and Lab-AO (e.g., \texttt{LABAO\_S1}). Where S1 being the name of a telescope and  names of other telescopes are S2, W1, W2, E1 and E2. The display of real-time images from the cameras are built with GTK + XWindow system and they are fast and use very less CPU and memory usage.   

\subsection{Automatic alignment of Tel-AO and Lab-AO for beam drifts}
The Lab-AO systems are connected to the Tel-AO systems using a metrology system. A dichroic at the telescope reflects some visible wavelengths of starlight into the Tel-AO WFS and also sends the beacon light into the Lab-AO WFS. It is very critical for the beacon light to follow the same path as the starlight. 

An automatic alignment software tracks telescope pupil and focus drifts in both the Tel-AO and Lab-AO WFS that can be caused by beacon position drifts and by the temperature changes during the course of observing. These pupil and focus drifts are corrected with several actuators, including M1, M2, M7, and M10 mirrors of the telescope as the Tel-AO's alignment is strongly coupled to the alignment of the telescope itself. The beacon alignment is strongly coupled to the quality of the Coudé path alignment. The AO system has several automatic alignment controls, and a few are listed below:

\begin{itemize}
\item Align the focus of the WFS feed parabola.
\item Align the beacon to the WFS.
\item Focusing the telescope sending starlight to the WFS. 
\item Align beacon focus.
\item Align of M7 of the telescope.
\item Align of M10 of the telescope.
\end{itemize}

\begin{figure} [h!]
\centering
\includegraphics[width=0.7\textwidth]{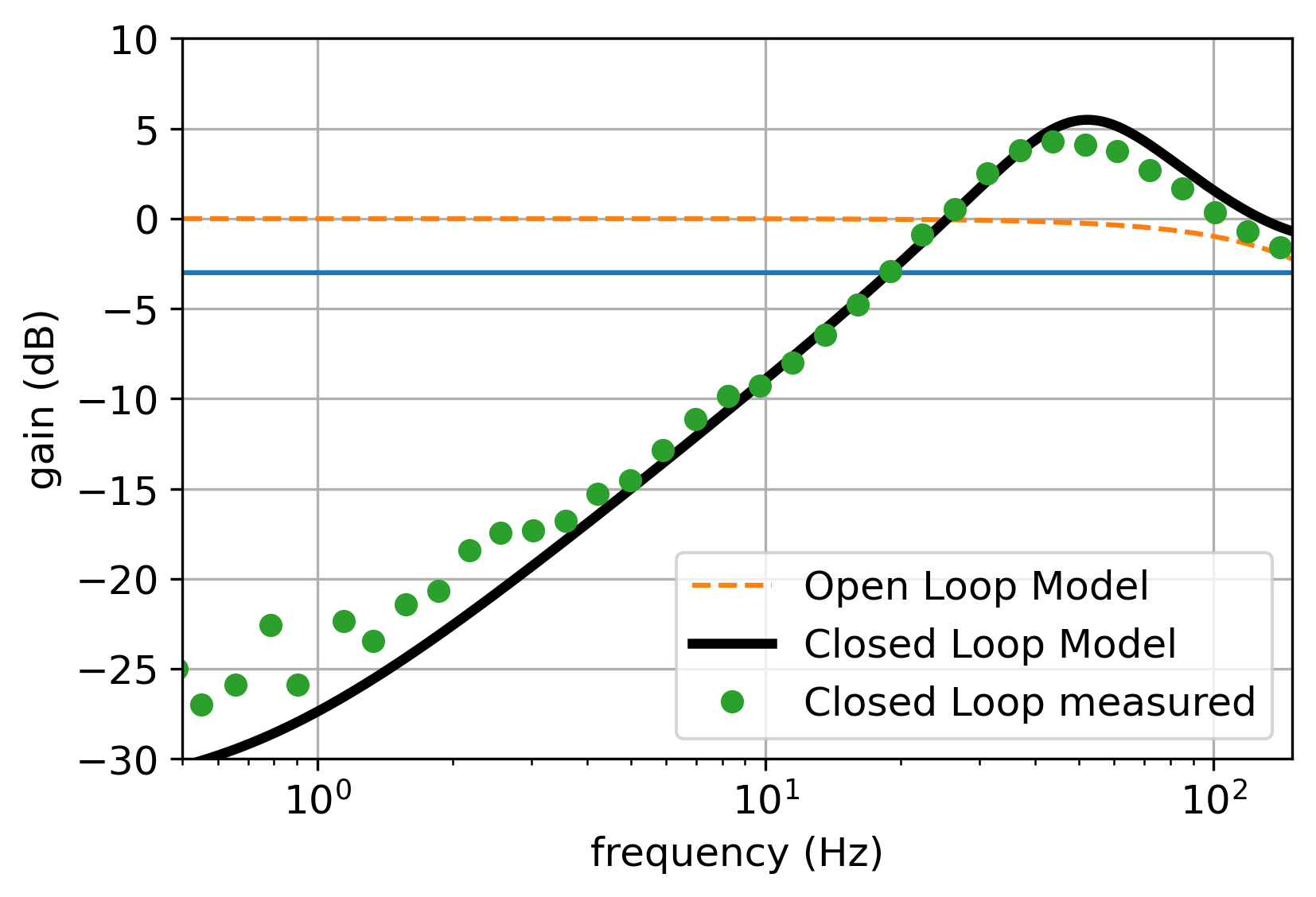}
\caption[example]{ \label{fig:telao_model} 
Rejection transfer function measured using the telescope beacon as the source of light.  The green dotted line is the measured closed-loop transfer function. The dashed line is the modeled open-loop transfer function. The thick black line is the  modeled closed-loop transfer function. This model uses the total delay time $T_{\rm D}$ is used. The green line cross the measured -3dB closed-loop bandwidth at frequency of 19Hz.  }
\end{figure} 

\subsection{The Tel-AO performance}
Figure~\ref{fig:telao_model} shows the performance of the Tel-AO for an integration time of 2ms. For this integration, the camera operates at a 440Hz measured frame rate and has a -3dB closed-loop bandwidth of 19Hz.  For comparison, this is almost similar performance to the adaptive optics system of the Auxiliary Telescopes of the VLTI (NAOMI)\cite{Woillez2019A&A...629A..41W} that runs at 500Hz and has -3dB closed-loop bandwidth of 18Hz. 

To verify the system performance, we have modeled the transfer function by putting the total delays of the control system, $T_{\rm D}$ (see Figure~\ref{fig:telao_model}). The total cycle time is the sum of the integration (2ms) and frame-transfer and readout time (1.8ms) that is $T_{\rm C}=3.8$ms for 2ms integration. The computational time for reading raw images from the instrument shared memory, computation of slopes, and computation of DM commands from slopes is $T_{\rm RTC}=0.7$ms. The total delay is $T_{\rm D} = T_{\rm C} + T_{\rm RTC}$.

\subsection{The Lab-AO performance}
The Lab-AO systems have been used since 2014 to measure a good ``default flat" of the sky close to the target star, and this has proved to be very helpful in coupling the light into the single-mode fibers used in many beam combiners\cite{Martinod2018A&A...618A.153M}.  We yet to measure the Tel-AO and Lab-AO's final performance in terms of sensitivity due to the COVID-19 and 2020 Bobcat Fire, but these systems already show increased throughput more than a magnitude.

\section{Summary and future plans}
We here report the CHARA Array adaptive optics systems from the software perspective -- highlighting operational complexity, solutions, and performance. An adaptive optics system's success depends on a lot of  ``end-to-end" closed-loop testing with its calibration source and the on-sky. We have tested all the subsystems, including simultaneous operation of  Tel-AO and Lab-AO since 2018 in various observing conditions. The Tel-AO's -3dB closed-loop transfer function bandwidth is measured at 19Hz, and this performance is almost similar to NAOMI instrument\cite{Woillez2019A&A...629A..41W}, which also uses the same model Andor 897 camera. The on-sky science observations with AOs are routine since the beginning of 2020. The on-sky tests are very promising, with more than a magnitude sensitivity improvement. We calculate more than 50\% Strehl ratio achieved in the H and K-bands. These results push observations of faint young stellar object disks and active galactic nuclei, and other interesting systems. The AO also push the observations in worse seeing conditions, which were not possible without the adaptive optics system before. The sensitivity improvements will be enjoyed by beam combines operating in the H and K-bands, for instance, MIRC-X\cite{Anugu2020AJ....160..158A}, CLASSIC/CLIMB\cite{ten2013} and upcoming MYSTIC\cite{Monnier2018}. 

The adaptive optics systems are used not only for improving the throughput but also for other science cases:
\begin{itemize}
    \item The Tel-AO and Lab-AO WFS pupil images can also be used to track the pupil between the telescope and the laboratory separated by many hundreds of meters of the optical path using starlight. This information is important for the estimation of accurate astrometry (e.g., MIRC-X astrometric mode\cite{Gardner2020}) by computing pupil shifts projected on the telescope space\cite{Anugu2018} during the course of observing. The Tel-AO and Lab-AO images from the telemetry will be used to measure pupil shifts in the post-processing.
   \item The Telescope DMs will be used to take sky thermal backgrounds for the upcoming MYSTIC\cite{Monnier2018} instrument by exploiting the large stroke of the ALPAO deformable mirrors.
\end{itemize}

We had to return two deformable mirrors to the ALPAO company for repairs. One for an upgrade and another to replace a bad actuator. They are expected to arrive in 2021 and will be installed once the COVID-19 situation is under control. During the first days of commissioning, the time taken to slew to a new target, acquisition, and locking all AO systems was almost double the slew time without the adaptive optics systems. However, since then, we have gained experience in our operations and reduced the total acquisition times. We are still learning and improving the better strategies for the interplay between the Tel-AO and Lab-AO systems and simplification of the operational aspects.

Our software is on the CHARA GitLab but private repository (https://gitlab.chara.gsu.edu/chara). If the reader is interested in accessing the software or discussing it, please feel free to contact the authors.

\acknowledgments 
This work is based upon observations obtained with the Georgia State University Center for High Angular Resolution Astronomy Array at Mount Wilson Observatory. The CHARA Array is supported by the National Science Foundation under Grant No. AST-1211929, AST-1411654, AST-1636624, and AST-1715788. Institutional support has been provided from the GSU College of Arts and Sciences and the GSU Office of the Vice President for Research and Economic Development.

\bibliography{report} 
\bibliographystyle{main} 

\end{document}